\documentclass[10pt,twocolumn, conference]{IEEEtran}

\usepackage{amsmath}
\usepackage{amssymb}
\usepackage[dvips]{graphicx}
\usepackage{epsfig}

\newcommand{\x}{\mathbf{x}}
\newcommand{\p}{\mathbf{P}}

\begin{document}
\title{Training Optimization for Gauss-Markov Rayleigh Fading Channels}
\author{\authorblockN{Sami Akin \hspace{1.0cm}
Mustafa Cenk Gursoy}
\authorblockA{Department of Electrical Engineering\\
University of Nebraska-Lincoln\\ Lincoln, NE 68588\\ Email:
sakin1@bigred.unl.edu, gursoy@unl.edu}}
\date{}

\maketitle

\begin{abstract}\footnote{This work was supported in part by the NSF CAREER Grant
CCF-0546384.} In this paper, pilot-assisted transmission over
Gauss-Markov Rayleigh fading channels is considered. A simple
scenario, where a single pilot signal is transmitted every $T$
symbols and $T-1$ data symbols are transmitted in between the
pilots, is studied. First, it is assumed that binary phase-shift
keying (BPSK) modulation is employed at the transmitter. With this
assumption, the training period, and data and training power
allocation are jointly optimized by maximizing an achievable rate
expression. Achievable rates and energy-per-bit requirements are
computed using the optimal training parameters. Secondly, a capacity
lower bound is obtained by considering the error in the estimate as
another source of additive Gaussian noise, and the training
parameters are optimized by maximizing this lower bound.
\end{abstract}
\section{Introduction}
One of the key characteristics of wireless communications that most
greatly impact system design and performance is the time-varying
nature of the channel conditions, experienced due to mobility and
changing physical environment. This has led mainly to three lines of
work in the performance analysis of wireless systems. A considerable
amount of effort has been expended in the study of cases in which
the perfect channel state information (CSI) is assumed to be
available at either the receiver or the transmitter or both. With
the perfect CSI available at the receiver, the authors in
\cite{Ericson} and \cite{Lee} studied the capacity of fading
channels. The capacity of fading channels is also studied in
\cite{Goldsmith1} and \cite{Goldsmith2} with perfect CSI at both the
receiver and the transmitter. A second line of work has considered
fast fading conditions, and assumed that neither the receiver nor
the transmitter is aware of the channel conditions (see e.g.,
\cite{FaycalShamai}, \cite{Marzetta}, \cite{gursoy1}). On the other
hand, most practical wireless systems attempt to learn the channel
conditions but can only do so imperfectly. Hence, it is of great
interest to study the performance when only imperfect CSI is
available at the transmitter or the receiver. When the channel is
not known a priori, one technique that provides imperfect receiver
CSI is to employ pilot signals in the transmission to estimate the
channel.

Pilot-Assisted Transmission (PAT) multiplexes known training signals
with the data signals. These transmission strategies and pilot
symbols known at the receiver can be used for channel estimation,
receiver adaptation, and optimal decoding \cite{Lang}. One of the
early studies has been conducted by Cavers in \cite{Cavers},
\cite{Cavers2} where an analytical approach to the design of PATs is
presented. \cite{Srihari} has shown that the data rates are
maximized by periodically embedding pilot symbols into the data
stream. The amount, placement, and fraction of pilot symbols in the
data stream have considerable impact on the data rate. The more
pilot symbols are transmitted and the more power is allocated to the
pilot symbols, the better estimation quality we have, but the more
time for transmission of data is missed and the less power we have
for data symbols. Hassibi and Hochwald \cite{babak} has optimized
the power and duration of training signals by maximizing a capacity
lower bound in multiple-antenna Rayleigh block fading channels. An
overview of pilot-assisted wireless transmission techniques is
presented in \cite{Lang}.

In \cite{AbouFaycalMedard}, considering adaptive coding of data
symbols without requiring feedback to the transmitter, Abou-Faycal
\emph{et al.} studied the data rates achieved with pilot symbol
assisted modulation (PSAM) over Gauss-Markov Rayleigh fading
channels. In this paper, the training period is optimized by
maximizing the achievable rates. The authors in \cite{AyahBdeir}
also considered pilot symbol-assisted transmission over Gauss-Markov
Rayleigh channels and analyzed the optimal power allocation among
data symbols while the pilot symbol has fixed power. They have shown
that the power distribution has a decreasing character with respect
to the distance to the last sent pilot, and that data power
adaptation improves the rates. The authors in \cite{Sen-Gur}
considered a similar setting and analyzed training power adaptation
but assumed that the power is uniformly distributed among data
symbols.

In this paper, considering that no prior channel knowledge is
available at the transmitter and the receiver, we focus on a
time-varying Rayleigh fading channel. The channel is modeled by a
Gauss-Markov model. Pilot symbols which are known by both the
transmitter and the receiver are transmitted with a period of $T$
symbols. In this setting, we seek to jointly optimize the training
period, training power, and data power allocation by maximizing
achievable rates.

\section{Channel Model}
We consider the following model in which a transmitter and a
receiver are connected by a time-varying Rayleigh fading channel,
\begin{equation} \label{channel_equ}
y_{k} = h_{k} x_{k} +n_{k} \quad k=1,2,3, \ldots
\end{equation}
where $y_{k}$ is the complex channel output, $x_{k}$ is the complex
channel input, $h_{k}$ and $n_{k}$ are the fading coefficient and
additive noise component, respectively. We assume that $h_{k}$ and
$n_{k}$ are independent zero mean circular complex Gaussian random
variables with variances $\sigma_{h}^2$ and $\sigma_{n}^2$,
respectively. It is further assumed that $x_{k}$ is independent of
$h_{k}$ and $n_{k}$.

While the additive noise samples $\{n_k\}$ are assumed to form an
independent and identically distributed (i.i.d.) sequence, the
fading process is modeled as a first-order Gauss-Markov process,
whose dynamics is described by
\begin{equation} \label{fading equ}
h_{k} = \alpha h_{k-1} + z_{k} \quad 0\leq\alpha\leq1, \quad k =
1,2,3,\ldots,
\end{equation}
where \{$z_{k}$\}'s are i.i.d. circular complex Gaussian variables
with zero mean and variance equal to (1-$\alpha^{2}) \sigma_{h}^2$.
In the above formulation, $\alpha$ is a parameter that controls the
rate of the channel variations between consecutive transmissions.
For instance, if $\alpha=1$, fading coefficients stay constant over
the duration of transmission, whereas, when $\alpha=0$, fading
coefficients are independent for each symbol. For bandwidths in the
10kHz range and Doppler spreads of the order of 100 Hz, typical
values for $\alpha$ are between 0.9 and 0.99
\cite{AbouFaycalMedard}.

\section{Pilot Symbol-Assisted Transmission}
We consider pilot-assisted transmission where periodically embedded
pilot symbols, known by both the sender and the receiver, are used
to estimate the fading coefficients of the channel thereby enabling
us to track the time-varying channel. We assume the simple scenario
where a single pilot symbol is transmitted every $T$ symbols while
$T-1$ data symbols are transmitted in between the pilot symbols. The
following average power constraint,
\begin{equation}\label{averagepower}
\frac{1}{T}\sum_{k=lT}^{(l+1)T-1}E\left[|x_{k}|^2\right]\leq P\quad
l = 0,1,2,\ldots,
\end{equation}
is imposed on the input. Therefore, the total average power
allocated to pilot and data transmission over a duration of $T$
symbols is limited by $PT$.

Communication takes place in two phases. In the training phase, the
pilot signal is sent and the channel output is given by
\begin{equation}\label{channeloutput}
y_{lT}=h_{lT}\sqrt{P_{t}}+n_{lT} \quad l=0,1,2,3, \ldots
\end{equation}where $P_{t}$ is the power allocated to the pilot
symbol. The fading coefficients are estimated via {MMSE} estimation,
which provides the following estimate:
\begin{equation} \label{eq:no55}
\widehat{h}_{lT}= \frac{\sqrt{P_{t}}\sigma_{h}^2}{P_{t}\sigma_{h}^2
+ \sigma_{n}^2} y_{lT}.
\end{equation}
Following the transmission of the training symbol, data transmission
phase starts and $T-1$ data symbols are sent. Since a single pilot
symbol is transmitted, the estimates of the fading coefficients in
the data transmission phase are obtained as follows:
\begin{equation} \label{eq:no6}
\widehat{h}_{k}=\frac{\sqrt{P_{t}}\sigma_{h}^2}{P_{t}\sigma_{h}^2 +
\sigma_{n}^2} \,\,\alpha^{k-lT}y_{lT} \quad lT  < k \le (l+1)T - 1.
\end{equation}Now, we can express the fading coefficients as
\begin{gather}
h_k = \widehat{h}_k + \widetilde{h}_k
\end{gather}
where $\widetilde{h}_k$ is the estimation error. Consequently, the
input-output relationship in the data transmission phase can be
written as
\begin{equation} \label{eq:channel2}
y_{k} = \widehat{h}_{k}x_{k}+\widetilde{h}_kx_{k}+n_{k} \quad lT  <
k \le (l+1)T - 1.
\end{equation}
Note that $\widehat{h}_k$ and $\widetilde{h}_k$ for $lT <k<(l+1)T$
are uncorrelated zero-mean circularly symmetric complex Gaussian
random variables with variances
\begin{gather} \label{eq:no7}
\sigma_{\widehat{h}_k}^2= \frac{P_{t}\sigma_{h}^4}{P_{t}\sigma_{h}^2
+ \sigma_{n}^2}(\alpha^{k-lT})^{2},
\\ \intertext{and}
\sigma_{\widetilde{h}_k}^2=\sigma_{h}^2-
\frac{P_{t}\sigma_{h}^4}{P_{t}\sigma_{h}^2 + \sigma_{n}^2}
(\alpha^{k-lT})^{2}, \label{eq:errorvar}
\end{gather}
respectively.

\section{Optimal Power Distribution and Training Period for BPSK Signals}

\subsection{Problem Formulation}

In this section, we consider that binary phase-shift keying (BPSK)
is employed at the transmitter to send the information. Since our
main goal is to optimize the training parameters and identify the
optimal power allocation, BPSK signaling is adopted due to its
simplicity. In the $k^{th}$ symbol interval, the BPSK signal can be
represented by two equiprobable points located at
$x_{k,1}=\sqrt{P_{d,k}}$ and $x_{k,2}=-\sqrt{P_{d,k}}$ on the
constellation map. Note that $P_{d,k}$ is the average power of the
BPSK signal in the $k^{th}$ symbol interval. In this interval, the
input-output mutual information conditioned on the value $y_{lT}$ is
given by
\begin{align} \label{mutualinfo}
&I_{k}(x_{k};y_{k}|y_{lT} = y_{lT}^*) = \nonumber\\ &=
\frac{1}{2}\int p_{y_{k}|x_{k}}(y|x_{k,1})\log
\frac{p_{y_{k}|x_{k}}(y|x_{k,1}))}{p_{y_{k}}(y)}dy \nonumber \\
& + \frac{1}{2}\int p_{y_{k}|x_{k}}(y|x_{k,2})\log
\frac{p_{y_{k}|x_{k}}(y|x_{k,2})}{p_{y_{k}}(y)}dy
\end{align}
where
\begin{gather*}
p_{Y_{k}|X_{k}}(y_{k}|x_{k})=
\frac{1}{\pi(\sigma_{\tilde{h}_k}^2|x_{k}|^2+\sigma_{n}^2)}
\exp\left(\frac{-|y_{k}-\widehat{h}_{k}x_{k}|^2}{\sigma_{\tilde{h}_k}^2|x_{k}|^2+\sigma_{n}^2}
\right) \\
\intertext{and} p_{y_k}(y_k) = \frac{1}{2}
p_{y_{k}|x_{k}}(y|x_{k,1}) + \frac{1}{2}p_{y_{k}|x_{k}}(y|x_{k,2}).
\end{gather*}
We consider the following achievable rate expression, which acts as
a lower bound to the channel capacity:
\begin{align}\label{rate}
I_{L}\left(T,P_{t},\mathbf{P}_{d} \right)&=E\left[\frac{1}{T}\sum_{k
= lT+1}^{(l+1)T-1} I_{k}(x_{k};y_{k}|y_{lT}=y_{lT}^*) \right] \\ &=
\frac{1}{T}\sum_{k=lT+1}^{(l+1)T-1}E\left[I_{k}(x_{k};y_{k}|y_{lT}=y_{lT}^*)\right]
\end{align}
where the expectation is with respect to $y_{lT}$, and $y_{lT}^*$ is
a realization of the random variable $y_{lT}$. Note that the
achievable rate is expressed as a function of the training period,
$T$; power of the pilot signal, $P_{t}$; and the power allocated to
$T-1$ data symbols transmitted in between the pilot symbols, which
is described by the following vector
\begin{align}
\mathbf{P}_d=\left[P_{d,1}, P_{d,2},...,P_{d,T-1} \right].
\end{align}
Our goal is to solve the joint optimization problem
\begin{gather} \label{eq:optprob}
(T^*, P_t^*, \mathbf{P}_d^*) = \arg \max_{\substack{T, P_t,
\mathbf{P}_d \\ P_t + \sum_{k = 1}^{T-1} P_{d,k} \le PT}} I_L(T,
P_t, \mathbf{P}_d)
\end{gather}
and obtain the optimal training period, and optimal data and pilot
power allocations. Since it is unlikely to reach to closed-form
solutions, we have employed numerical tools to solve
(\ref{eq:optprob}).

\subsection{Numerical Results}

\begin{figure}
\begin{center}
\includegraphics[width = 0.5\textwidth]{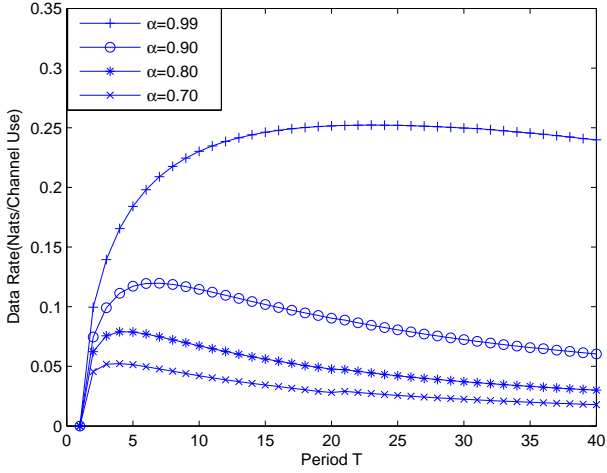}
\caption{Achievable data rates vs. training period $T$ for
$\alpha=0.99$,$0.90$,$0.80$, and $0.70$.
$SNR=\frac{P}{\sigma_{n}^{2}}=0dB$} \label{fig:fig1}
\end{center}
\end{figure}
\begin{figure}
\begin{center}
\includegraphics[width = 0.5\textwidth]{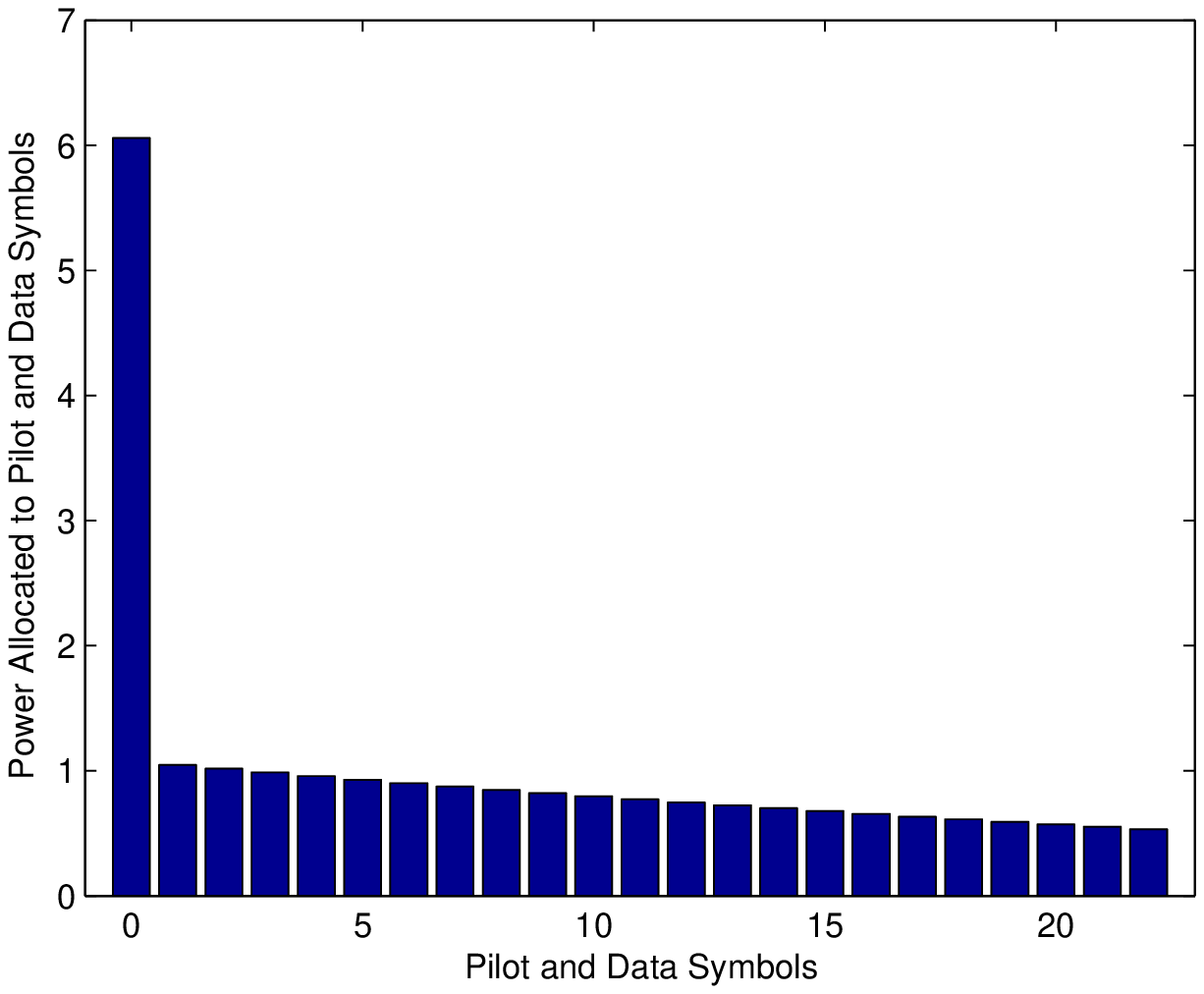}
\caption{Optimal power distribution among the pilot and data symbols
when $\alpha=0.99$ and SNR=0dB. The optimal period is $T=23$.}
\label{fig:fig2}
\end{center}
\end{figure}
\begin{figure}
\begin{center}
\includegraphics[width = 0.5\textwidth]{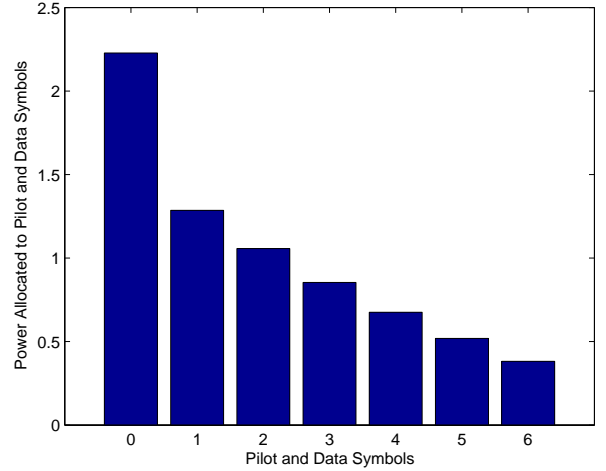}
\caption{Optimal power distribution among the pilot and data symbols
when $\alpha=0.90$ and SNR=0dB. The optimal period is $T=7$.}
\label{fig:fig3}
\end{center}
\end{figure}
\begin{figure}
\begin{center}
\includegraphics[width = 0.5\textwidth]{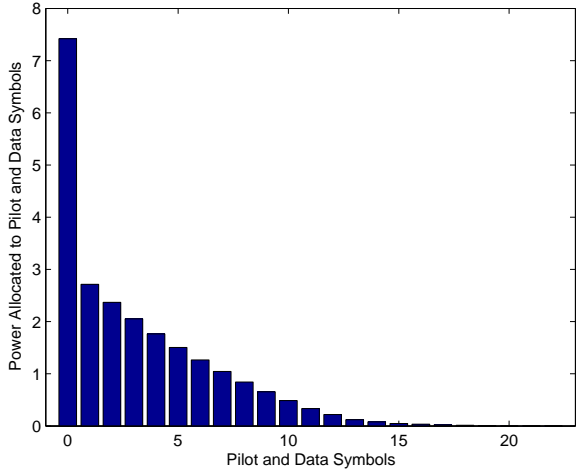}
\caption{Optimal power distribution among the pilot and data symbols
when $\alpha=0.90$ and SNR=0dB. The optimal period is $T=23$.}
\label{fig:fig4}
\end{center}
\end{figure}

In this section, we summarize the numerical results. Figure
\ref{fig:fig1} plots the data rates achieved with optimal power
allocations as a function of the training period for different
values of $\alpha$. The power level is kept fixed at $P = \sigma_n^2
= 1$. It is observed that the optimal values of the training period,
$T$, are 23, 7, 4, and 4 for $\alpha=0.99, 0.90, 0.80$, and $0.70$,
respectively. Note that the optimal $T$ and optimal data rate are
decreasing with the decreasing $\alpha$. This is expected because
the faster the channel changes, the more frequently the pilot
symbols should be sent. This consequently reduces the data rates
which are already adversely affected by the fast changing and
imperfectly known channel conditions. Figures \ref{fig:fig2} and
\ref{fig:fig3} are the bar graphs providing the optimal training and
data power allocation when the training period is at its optimal
value. In the graphs, the first bar corresponds to the power of the
training symbol while the remaining bars provide the power levels of
the data symbols. We immediately observe from both figures that the
data symbols, which are farther away from the pilot symbol, are
allocated less power since channel gets noisier for these symbols
due to poorer channel estimates. Moreover, comparing Fig.
\ref{fig:fig2} and Fig. \ref{fig:fig3}, we see that having a longer
training period enables us to put more power on the pilot signal and
therefore have better channel estimates. We also note that if
$\alpha$ is small as in Fig. \ref{fig:fig3}, the power of the data
symbols decreases faster as they move away from the pilot symbol.
From these numerical results, it is evident that $\alpha$ greatly
affects the optimal power allocation and optimal $T$. Fig.
\ref{fig:fig4} gives the power distribution when $\alpha = 0.90$ and
$T = 23$. Note this value of the training period is suboptimal. The
inefficiency of this choice is apparent in the graph. Since the
channel is changing relatively fast and the quality of the channel
estimate deteriorates rather quickly, only half of the available
time slots are used for data transmission, leading to a considerable
loss in data rates.

\begin{figure}
\begin{center}
\includegraphics[width = 0.5\textwidth]{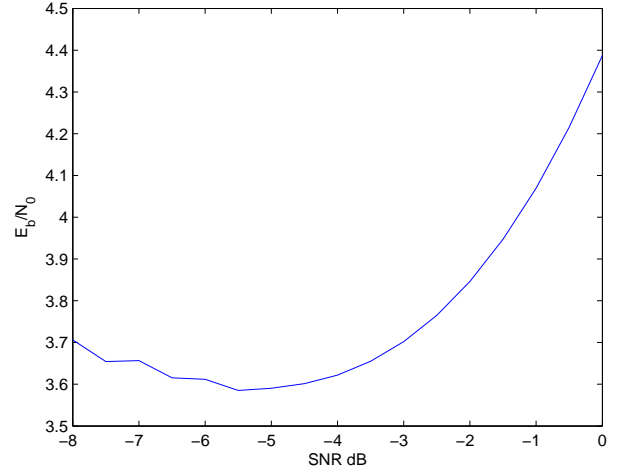}
\caption{Bit energy $\frac{E_{b}}{N_{0}}$ vs. SNR(dB) when
$\alpha=0.99$.} \label{fig:fig5}
\end{center}
\end{figure}

\begin{figure}
\begin{center}
\includegraphics[width = 0.5\textwidth]{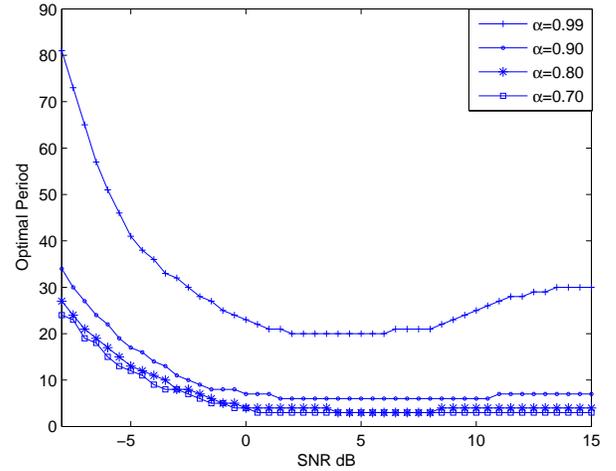}
\caption{Optimal training period T vs. SNR for
$\alpha=0.99,0.90,0.80,0.70$.} \label{fig:fig6}
\end{center}
\end{figure}

\begin{figure}
\begin{center}
\includegraphics[width = 0.5\textwidth]{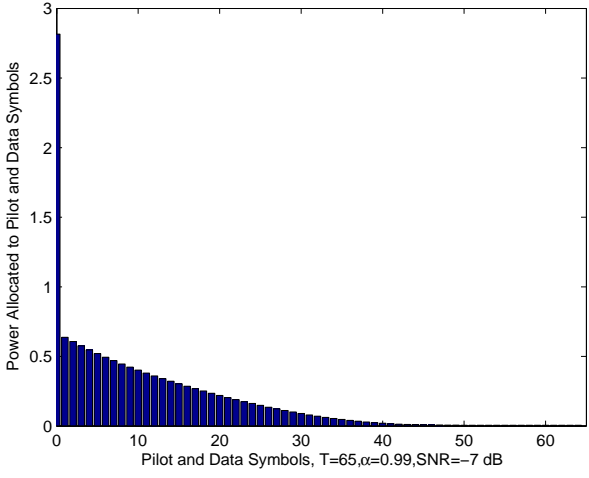}
\caption{Optimal power distribution for the pilot and data symbols
when $\alpha=0.99$ and SNR=-7dB. The optimal period is T=65.}
\label{fig:fig7}
\end{center}
\end{figure}

In systems with scarce energy resources, energy required to send one
information bit, rather than data rates, is a suitable metric to
measure the performance. The least amount of normalized bit energy
required for reliable communications is given by
\begin{gather} \label{eq:leastminbitenergy}
\frac{E_b}{N_0} = \frac{\text{SNR}}{C(\text{SNR})}
\end{gather}
where $C(\text{SNR})$ is the channel capacity in bits/symbol. In our
setting, the bit energy values found from
\begin{equation}\label{maxcapacitySNR}
\frac{E_b}{N_0} = \frac{\text{SNR}}{I_L(T^*, P_t^*, \mathbf{P}_d^*)}
\end{equation}
provide an upper bound on the values obtained from
(\ref{eq:leastminbitenergy}), and also gives us indications on the
energy efficiency of the system. Fig. \ref{fig:fig5} plots the
required bit energy values as a function of the SNR. The bit energy
initially decreases as SNR decreases and achieves its minimum value
at approximately SNR$= -5.5$ dB below which the bit energy
requirement starts increasing. Hence, it is extremely energy
inefficient to operate below SNR$=-5.5$ dB. In general, one needs to
operate at low SNR levels for improved energy efficiency. From Fig.
\ref{fig:fig6}, which plots the optimal training period, $T$, as a
function of the SNR, we observe that $T$ increases as SNR decreases.
Hence, training is performed less frequently in the low SNR regime.
Fig. \ref{fig:fig7} provides the pilot and data power allocation
when SNR = $-$7dB, $\alpha = 0.99$, and $T = 65$. It is interesting
to note that although $T$ is large, a considerable portion of the
available time slots are not being used for transmission. This
approach enables the system to put more power on the pilot symbol
and nearby data symbols. Hence, although the system trains and
transmits less frequently, a more accurate channel estimate is
obtained and used in return.

\section{Low Complexity Training Optimization}
\begin{figure}
\begin{center}
\includegraphics[width = 0.5\textwidth]{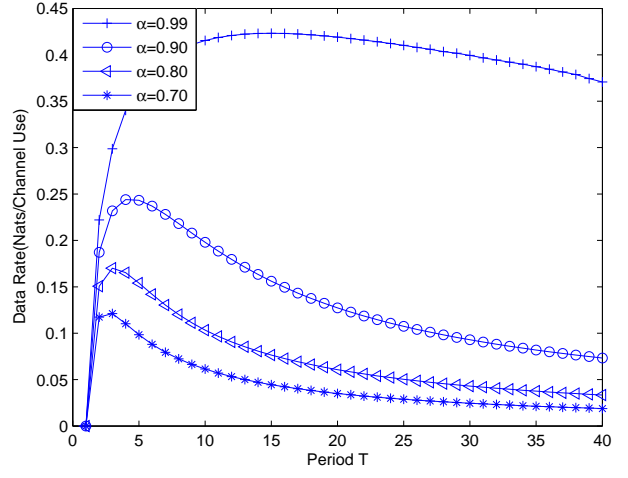}
\caption{Achievable data rates vs. training period T for
$\alpha=0.99,0.90,0.80$, and 0.70. SNR=5dB} \label{fig:fig8}
\end{center}
\end{figure}

Recall that the input-output relationship in the data transmission
phase is given by\footnote{It is assumed that a single pilot signal
is transmitted at $k = 0$. }
\begin{equation} \label{eq:channel3}
y_k = \widehat{h}_kx_k + \widetilde{h}_kx_k + n_k \quad k =
1,2,\ldots, T-1.
\end{equation}
In the preceding section, we fixed the modulation format and
computed the input-output mutual information achieved in the channel
(\ref{eq:channel3}). In this section, we pursue another approach
akin to that in \cite{babak}. We treat the error in the channel
estimate as another source of additive noise and assume that
\begin{equation} \label{eq:newnoise}
w_k =\widetilde{h}_kx_k + n_k
\end{equation}
is zero-mean Gaussian noise with variance
\begin{equation}\label{variancenew}
\sigma_{w_{k}}^2 = \sigma_{\widetilde{h}_k}^2P_{d,k}+\sigma_{n}^2.
\end{equation}
where $P_{d,k} = E[|x_k|^2]$ is the average power of the symbol
$x_k$ and $\sigma_{\widetilde{h}_k}^2$ is given in
(\ref{eq:errorvar}). Since the Gaussian noise is the worst case
noise \cite{babak}, the capacity of the channel
\begin{gather} \label{eq:worstcasechannel}
y_k = \widehat{h}_kx_k + w_k \quad k = 1,2,\ldots
\end{gather}
is a lower bound to the capacity of the channel given in
(\ref{eq:channel3}). An achievable rate expression for channel
(\ref{eq:worstcasechannel}) is
\begin{align}
I_{\text{worst}} &= \max_{T, P_t} \!\!\!\max_{\substack{ \x \\
E[|\x|^2] \le PT - P_t}} \!\!\frac{1}{T}\sum_{k=1}^{T-1}
I_{k}(x_{k};y_{k}|\widehat{h}_k) \label{eq:worstcase1}
\\
&=\max_{T, P_t} \!\!\!\max_{\substack{\p_d \\ P_{d,k} \ge 0 \,\, \forall k\\
\sum_{k = 1}^{T-1} P_{d,k} \le PT-P_t}}
\!\!\!\frac{1}{T}\sum_{k=1}^{T-1}\max_{\substack{x_k
\\ E[|x_k|^2] \le
P_{d,k}}}\!\!\!I_{k}(x_{k};y_{k}|\widehat{h}_k)
\label{eq:worstcase2}
\\
&=\!\!\!\!\!\!\!\!\!\!\max_{\substack{T, P_t, \p_d \\ \sum_{k =
1}^{T-1} P_t+P_{d,k} \le PT}} \!\!\!\frac{1}{T}\sum_{k=1}^{T-1}
E\left[ \log\left(1 +
\frac{\sigma_{\widehat{h}_k}^2P_{d,k}}{\sigma_{\widetilde{h}_k}^2P_{d,k}+\sigma_{n}^2}|\xi|^2\right)\right].
\label{eq:worstcase3}
\end{align}
In (\ref{eq:worstcase1}), $\x = (x_1, x_2, \ldots, x_{T-1} )$
denotes the vector of $T-1$ input symbols, and the inner
maximization is over the space of joint distribution functions of
$\x$. (\ref{eq:worstcase2}) is obtained by observing that once the
data power distribution is fixed, the maximization over the joint
distribution can be broken down into separate maximization problems
over marginal distributions. (\ref{eq:worstcase3}) follows from the
fact that Gaussian input maximizes the mutual information $I(x_k ;
y_k|\widehat{h}_k)$ when the channel in consideration is
(\ref{eq:worstcasechannel}). Note that in (\ref{eq:worstcase3}),
$\xi$ is a zero mean, unit variance, circular complex Gaussian
random variable, and the expectation is with respect to $\xi$. We
can again numerically solve the above optimization and Fig.
\ref{fig:fig8} plots the achievable data rates with optimal power
allocation as a function of $T$ for different values of $\alpha$
when SNR=5dB.
An even simpler optimization problem results if we seek to optimize
the upper bound
\begin{align}
\frac{1}{T}\sum_{k=1}^{T-1} &E\left[ \log\left(1 +
\frac{\sigma_{\widehat{h}_k}^2P_{d,k}}{\sigma_{\widetilde{h}_k}^2P_{d,k}+\sigma_{n}^2}|\xi|^2\right)\right]
\\ &\le \frac{1}{T}\sum_{k=1}^{T-1} \log\left(1 +
\frac{\sigma_{\widehat{h}_k}^2P_{d,k}}{\sigma_{\widetilde{h}_k}^2P_{d,k}+\sigma_{n}^2}\right),
\end{align}
which is obtained by using the Jensen's inequality and noting that
$E[|\xi|^2] = 1$. In this case, the optimization problem becomes
\begin{align} \label{capacityworst}
&\max_{\substack{T, P_t, \p_d \\ \sum_{k = 1}^{T-1} P_t+P_{d,k} \le
PT}} \!\!\! \frac{1}{T}\sum_{k=1}^{T-1} \log\left(1 +
\frac{\sigma_{\widehat{h}_k}^2P_{d,k}}{\sigma_{\widetilde{h}_k}^2P_{d,k}+\sigma_{n}^2}\right)
\\
&=\max_{\substack{T, P_t, \p_d \\ \sum_{k = 1}^{T-1} P_t+P_{d,k} \le
PT}} \!\!\! \frac{1}{T} \log\left( \prod_{k = 1}^{T-1}\left(1 +
\frac{\sigma_{\widehat{h}_k}^2P_{d,k}}{\sigma_{\widetilde{h}_k}^2P_{d,k}+\sigma_{n}^2}\right)\right).
\end{align}
Since logarithm is a monotonically increasing function, the optimal
training and data power allocation for fixed $T$ can be found by
solving
\begin{align} \label{eq:capacitymax}
\max_{\substack{P_t, \p_d \\ \sum_{k = 1}^{T-1} P_t+P_{d,k} \le PT}}
\prod_{k = 1}^{T-1}\left(1 +
\frac{\sigma_{\widehat{h}_k}^2P_{d,k}}{\sigma_{\widetilde{h}_k}^2P_{d,k}+\sigma_{n}^2}\right).
\end{align}
\begin{figure}
\begin{center}
\includegraphics[width = 0.5\textwidth]{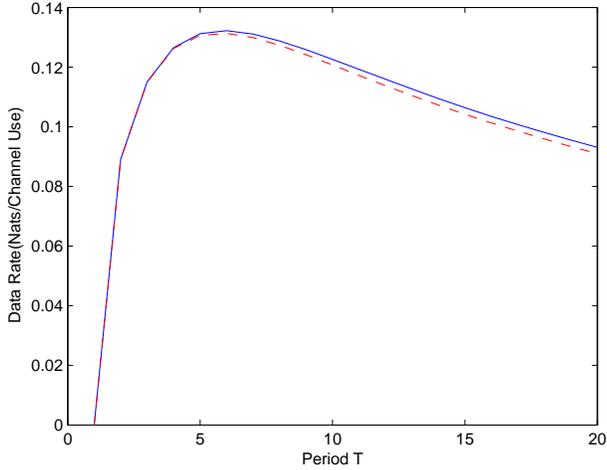}
\caption{Achievable data rates for BPSK signals vs.training period T
for $\alpha=0.90$. SNR=0dB.} \label{fig:fig9}
\end{center}
\end{figure}
\begin{figure}
\begin{center}
\includegraphics[width = 0.5\textwidth]{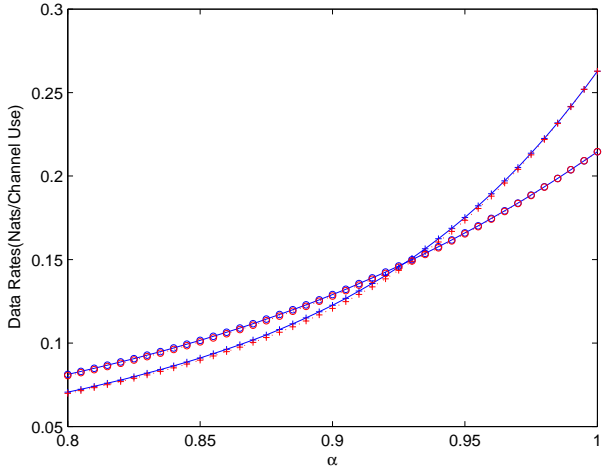}
\caption{Achievable data rates for BPSK signals vs. $\alpha$ for $T
= 6$ and $10$. SNR = 0 dB. "+ and solid line" and "+ and dotted
line" are plotting rates achieved with power allocation from
(\ref{eq:capacitymax}) and (\ref{eq:optprob}), respectively, when $T
=10$. "o and solid line" and "o and dotted line" are plotting rates
achieved with power allocation from (\ref{eq:capacitymax}) and
(\ref{eq:optprob}), respectively, when $T =6$.} \label{fig:fig10}
\end{center}
\end{figure}
It is very interesting to note that the optimal power distribution
found by solving (\ref{eq:capacitymax}) is very similar to that
obtained from (\ref{eq:optprob}) where BPSK signals are considered.
Figure \ref{fig:fig9} plots the achievable data rates as a function
of training period when BPSK signals are employed for transmission.
Hence, the data rates are computed using (\ref{rate}). In the
figure, the solid line shows the data rates achieved with power
distribution found from (\ref{eq:optprob}) while the dashed line
corresponds to rates achieved with power allocation obtained from
(\ref{eq:capacitymax}). Note that both curves are very close and the
training period is maximized at approximately the same value.

Fig. \ref{fig:fig10} plots the achievable rates for BPSK signals as
a function of the parameter $\alpha$ for $T = 6$ and $10$. The power
distribution is again obtained from both (\ref{eq:capacitymax}) and
(\ref{eq:optprob}). We again recognize that the loss in data rates
is negligible when (\ref{eq:capacitymax}) is used to find the power
allocation.

\section{Conclusion}

We have studied the problem of training optimization in
pilot-assisted wireless transmissions over Gauss-Markov Rayleigh
fading channels. We have considered a simple scenario where a single
pilot is transmitted every $T$ symbols for channel estimation and
$T-1$ data symbols are transmitted in between the pilot symbols.
MMSE estimation is employed to estimate the channel. We have jointly
optimized the training period, $T$, and data and training power
distributions by maximizing achievable rate expressions. We have
provided numerical results showing the optimal parameters, power
distributions, and maximized achievable rates. We have also studied
the energy efficiency of pilot-assisted transmissions by considering
the energy-per-bit requirements.

\end{document}